# Magneto and ferroelectric phase transitions in HoMn$_2$O$_5$ monocrystals

I. RADULOV, V. LOVCHINOV AND M. DASZKIEWICZ[a)]
*Institute of Solid State Physics, Bulgarian Academy of Sciences, Sofia, Bulgaria*
a) *Institute for Structural Research and Low Temperatures, Polish Academy of Sciences, Wroclaw, Poland*
**Corresponding author's e-mail: radulov@issp.bas.bg**

From the physical point of view multiferroics present an extremely interesting class of systems and problems. These are essentially of two kinds. One is what are the microscopic conditions, and sometimes constrains, which determine the possibility to combine in one system both magnetic and ferroelectric properties. This turned out to be a quite nontrivial question, and usually, in conventional systems, these two phenomena tend to exclude one another. Why it is the case is an important and still not completely resolved issue. In the present article we report our results from magnetic properties measurements on HoMn$_2$O$_5$ with short discussion about it possible origin.

## 1. Introduction

The researcher's interest to manganese oxides compounds with perovskite structure grew sharply after the discovery of the giant magnetoresistance effect in these compounds. For some of them, this effect was observed also near room temperatures, which was an indication for possible practical applications in different sensor devices. The heightened scientific activity in this area leaded to the discovery of more interesting and complicated effects in REMnO$_3$ and REMn$_2$O$_5$ (RE = rare earth) compounds reported recently (Colossal magnetodielectric effects in DyMn$_2$O$_5$ [1], magnetic field induced polarization and depolarization [2], Colossal magnetostriction effect in HoMn$_2$O$_5$ [3] etc.). All these effects results from the coexistence and mutual interference of long range orders, such as magnetic, ferromagnetic, antiferromagnetic, ferroelectric etc. In spite of all experimental investigations, the physical origin of the large magneto-electric coupling and the ferroelectricity arising at the magnetic lock-in transitions is not yet understood.

## 2. Sample preparation and properties

HoMn$_2$O$_5$ monocrystals were grown as described in [3]. HoMn$_2$O$_5$ (Ho$^{3+}$Mn$^{3+}$Mn$^{4+}$O$_5^{2+}$) is magnetic ferroelectric which undergoes the ferroelectric transition at $T'_N$ ~22 K below the antiferromagnetic transition temperature $T_N$ = 43 K.

The ferroelectric (FE) state exhibits canted antiferroelectric displacements of the Mn$^{3+}$ ions. These displacements are expected to lift the magnetic degeneracy by lowering the crystal symmetry to $P_{b21m}$, thus stabilizing the FE state via the magnetic Jahn-Teller effect. In the last years, a number of attempts to detect and to characterize the structural distortions in REMn$_2$O$_5$ by neutron scattering experiments, have been resulted inadequate due to the limited resolution of such experiments (TbMn$_2$O$_5$ and DyMn$_2$O$_5$ [4, 5]), or failed completely (HoMn$_2$O$_5$ and YMn$_2$O$_5$ [5 – 7]). Another possible way to resolve the structural distortions in REMn$_2$O$_5$ is to investigate the coupling between magnetic, dielectric, and lattice degrees of freedom.

The structure of HoMn$_2$O$_5$ consists of edge-sharing Mn$^{4+}$O$_6$ octahedra forming chains along the *c* axis crosslinked via Mn$^{3+}$O$_5$ pyramidal units designating the space group D$^9_{2h}$ − *Pbam* at room temperature and leading to five independent nearest neighbors (NN) magnetic interactions. The various connectivity of the Mn$^{4+}$O$_6$ octahedra and Mn$^{3+}$O$_5$ pyramids lead to five different NN exchange interactions. The Mn$^{4+}$O$_6$ octahedra share at the Ho and the Mn$^{3+}$ layers common O2-O2 and O3-O3 edges respectively. The resulting exchange interactions are shown on Fig. 1 as J1 and J2. The J5 exchange interaction results from every two Mn$^{3+}$O$_5$ pyramids connected to each other at the edge of their bases O1-O1. The connection of the Mn$^{3+}$O$_5$ pyramids to the Mn$^{4+}$O$_6$ octahedra through the apex oxygen O3 or base oxygen O4 determinate exchange interactions J4 and J3 correspondingly, which are relevant to the ***ab*** plane, but the competition between the different exchange interaction is not confined to the ***ab*** plane. The weak superexchange associated with J2 is expected to support an AFM arrangement of the Mn$^{4+}$ spins in the edge-shared Mn$^{4+}$O$_6$ octahedra. On the other hand both, J3 and J4 exchange interactions, favour the FM alignment of the same $Mn^{4+}$ spins. The frustrated topology makes it impossible to satisfy all of the favorable interactions simultaneously. This way every $Mn^{4+}$ moment has one NN $Mn^{3+}$ moment in the **b** direction with the "wrong" sign. It appears that |J4| > |J3| and that J4 is always AFM. The AFM J4 exchange interaction forms zig-zag chains in which pairs of J4 exchange interaction are separated by an AFM J5 interaction. All this results the ubiquitous doubling of the **a** axis (as discussed bellow and shown on Fig. 5). The Mn-Mn separation across the shared edges is short enough for both the $Mn^{4+}O_6$ octahedra (J1/J2) and $Mn^{3+}O_5$ pyramids (J5) exchange interaction, so that the direct exchange has dominant contribution for these interactions [4, 5].

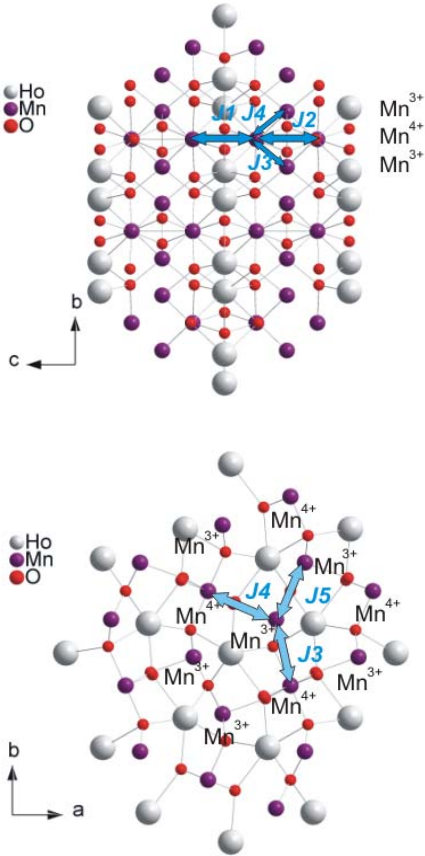

Fig. 1 Structure of $HoMn_2O_5$ with 5 resulting NN exchange interactions

## 3. Results and discussion

Mn and Ho ions in $HoMn_2O_5$ form three magnetic subsystems: $Ho^{3+}$, $Mn^{3+}$ and $Mn^{4+}$ with crystallographic position *4g*, *4h* and *4f* respectively. Magnetic interactions in the Mn subsystem are geometrically frustrated and $HoMn_2O_5$ shows complicated series of magnetic transitions involving the Mn and Ho ions on cooling below $T_N$ =43 K. The $Mn^{3+}$ ions have a decisive importance for the ferroelectric ordering of the $REMn_2O_5$ system [7].

The dielectric constant and polarization measurements displayed in Figs. 2 and 3, exhibit distinct anomalies at all magnetic transitions with the strongest peaks observed at the FE transitions. Our data provide striking evidence for the existence of extraordinarily large spin-lattice interactions in $HoMn_2O_5$. Further they prove that the magnetic and lattice degrees of freedom are strongly coupled and the simultaneous magnetoelectric (ME) and FE transitions at $T_C$ and $T'_N$ have to be considered as the phase change of one highly correlated system.

It is characteristic for all $REMn_2O_5$ compounds that the various magnetic phase changes are reflected in sharp and distinct anomalies of the dielectric constant, as shown in Fig.2. This is a clear indication for strong magneto-electric coupling due to large spin-lattice interactions.

Long-range antiferromagnetic (AFM) ordering of the $Mn^{3+}/Mn^{4+}$ spins occur at $T_N$ = 43 K. This transition into a high-temperature Néel phase is the common features for all $REMn_2O_5$, and is characterized by a two-component incommensurate magnetic modulation with wave vector $q$ = (qx, 0, qz). Subsequently the FE transition takes place at $T_C$ = 39 K slightly below $T_N$. The pure ferroelectric lock-in transition, observed at 39 K with $q$ = (0.5, 0, 0.25) is not influenced by magnetic fields. It has been assumed that the long-range magnetic ordering of $Mn^{3+}/Mn^{4+}$ induces the FE transition via an additional Jahn-Teller distortion of $Mn^{3+}$ ions [10].

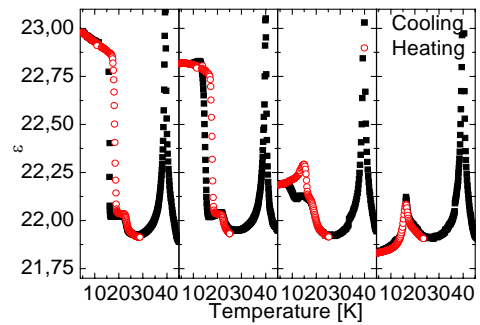

Fig.2 Temperature dependences of the dielectric constant along the **b** axis by sample cooling/ heating (filled square/open circle). Magnetic field is applied parallel to the **b** axis.

With further decreasing temperature, at $T'_N$, $HoMn_2O_5$ is characterized by the existence of another magnetic transition at $T'_N \sim 22$ K, at which commensurate AFM (c AFM) ordering becomes low temperatures incomemensurate (ic AFM) with a magnetic modulation $q \approx (0.5, 0, 0.3)$ [5–9]. The transition at $T'_N$ is accompanied by a significant decrease of the FE polarization and is often referred to as a second FE phase transition.

Under $T'_N$ $q$ remains unchanged (0.5, 0, 0.33) and the transition involves an increase in the ordered moments of the $Mn^{4+}/Mn^{3+}$ sublattice. Below 19 K a phase transition to canted AFM (CAFM I) takes place. It was found from our measurements that second CAFM- type ordering (CAFM II) of Ho ions occurs at $T_{N(Ho)}$ below 11 K.

Measurements at low magnetic fields show peculiarities around 11 K. In zero fields, the dielectric permittivity has hysteresis in the range 4.2 – 20 K, which becomes broader in higher fields, up to 12 T, where disappears. The last two transitions, at $T_{N(Ho)}$ and $T'_N$, change significantly their shape and place, depending on the intensity of the applied magnetic field. This is a clear indication for their magnetic origin. The phase diagram build from data received by dielectric constant measurements of HoMn$_2$O$_5$ is shown in Fig. 4 (left block).

polarization is reached at temperature of about 13 K, where a phase transition to CAFM (CAFM I) ordering take place. By further cooling the polarization of the sample increases fast and after a small change in the region of 10 K, which can be explained by $T_{N(Ho)}$ (CAFM II), remains near by constant down to 4.2 K. Also here a hysteresis in the range 4.2 K - 20 K and fields up to 12 T was observed, which disappears in higher fields. The phase diagram built from HoMn$_2$O$_5$ polarization measurements is shown on Fig. 4 (right block).

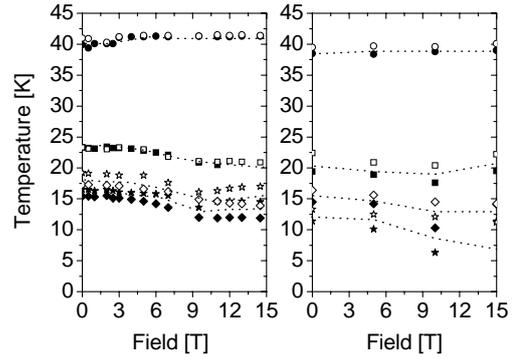

*Fig. 4 Phase diagrams based on the dielectric constant (left block) and polarization (right block) measurements. Open/ filled symbols means data obtained by sample cooling/ heating respectively.*

The spins of two $Mn^{3+}$ per unit cell are each frustrated with two neighboring $Mn^{4+}$ with the same spin direction. Reducing this frustration by moving the $Mn^{3+}$ away from the $Mn^{4+}$ generates a dipolar moment **P** (see Fig. 5) between the $Mn^{3+}$ and the surrounding oxygen ions. The **P$_a$**-components of this dipolar moments cancel out while the **P$_b$**-components add up to the macroscopic polarization and ferroelectricity along the *b* axis. The proposed displacement lowers the symmetry to the space group $P_{b21m}$. The AFM modulation along the *a* axis with qx=0.5 leads to the frustration and displacement of both $Mn^{3+}$ and the net polarization along the *b* axis.

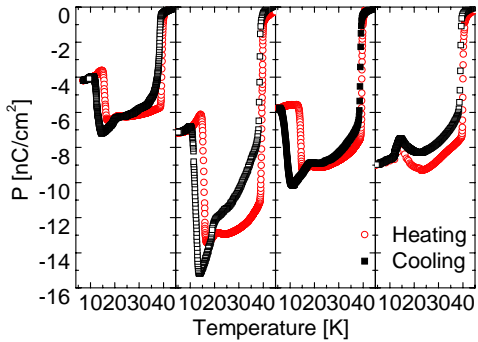

*Fig.3 Temperature dependences of the polarization along the b axis by sample cooling/ heating (filled square/open circle). Magnetic field is applied parallel to the b axes.*

The spontaneous polarization along the **b** axes appears by cooling the sample below $T_N = 43$ K. The spontaneous polarization remains negative for all fields in the range 4.2 – 50 K. HoMn$_2$O$_5$ undergoes a transition to a ferroelectric phase at $T_C \sim 39$ K, where the absolute value of the spontaneous polarization increases vastly. Further it changes slightly to temperatures about 20 K, where a new sharp change takes place. This change can be explained with the spontaneous reorientation of the magnetic lattice at $T'_N$ (c/ic AFM ordering, accompanied by changes in the dielectric properties). The maximum value of the

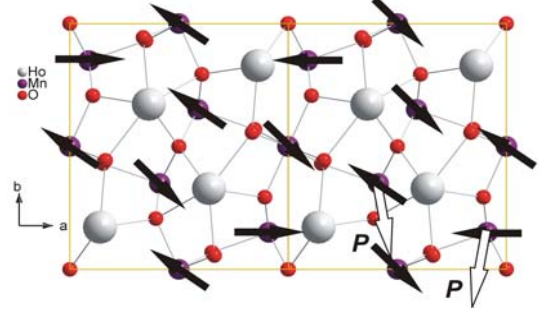

*Fig. 5 Spin and dipolar moment (black and white arrows) orientation in doubled HoMn$_2$O$_5$ unit cell.*

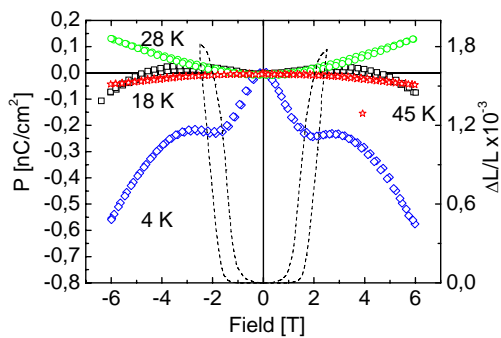

*Fig. 6 Field dependence of the polarization along the **b** axis at 4.2, 18, 28 and 45 K (diamond, square, circle and star). The magnetic field is applied parallel to the **b** axes.*

On Fig. 6 the polarization vs. field dependences of $HoMn_2O_5$ at different temperatures are presented. At 4.2 K a small peculiarity appears around 1.5 T. The magnetostriction of $HoMn_2O_5$ at 4.2 K (dash-line curve) also has a peculiarity at 1.5 T, where the giant magnetostriction effect appears. This can be an indication for a possible correlation between the field-induced polarization and magnetostriction. To confirm this correlation more detailed measurements in this temperature region are needed.

### 4. Conclusions

In the temperature range 43 − 4 K a consistency of magnetoelastoelectric phase transitions was observed. It was pointed out that all the ferroelectric phases are strongly tied to the antiferromagnetic Mn3+/Mn4+ spin structure, with the latter being dominated by the f–d exchange interaction. The appearance of ferroelectricity is a consequence of frustration between NN and NNN (next-nearest neighbour) $Mn^{4+}$ in the lattice. The frustration is lifted by Jahn–Teller distortion, and the associated reduction of symmetry allows the formation of a spontaneous polarization.

Considering the role of magnetic frustration to stabilize the ferroelectricity in $HoMn_2O_5$ there are interesting similarities to multiferroic $Ni_3V_2O_8$ and $TbMnO_3$. By other compounds it was shown that the transition sinusoidal to helical magnetic modulation can introduce a third order coupling giving rise to FE order [10]. On the other hand more detailed treatment shows that the existence of a spiral magnetic structure alone is not yet sufficient for FE: not all the spiral can lead to it. As shown in [11] FE can appear if the spin rotation axis **e** does not coincide with the wave vector of a spiral *Q*: the polarization **P** appears only if these two directions are different and it is proportional to the vector product of **e** and *Q*: **P**~*Q* x **e.**

However, it is not clear yet if the magnetic structure between $T_C$ and $T_N$ is sinusoidal and the transition into the FE phase follows the same mechanisms as in $Ni_3V_2O_8$ or $TbMnO_3$. Furthermore, the magnetic modulation in the FE phase of $HoMn_2O_5$ is commensurate whereas it is incommensurate in $Ni_3V_2O_8$ or $TbMnO_3$.

The strong magnetoelectric correlation is indicated by the observation that ordering of the Mn spins modifies the dielectric function, while ferroelectric ordering leaves an imprint on the magnetic susceptibility.

### 5. Acknowledgements

The work of I. Radulov is supported by NATO EAP.RIG.981824.